\documentclass[preprint,showpacs,preprintnumbers,prb,amsmath,amssymb]{revtex4}

\usepackage{graphicx}% Include figure files
\usepackage{dcolumn}% Align table columns on decimal point
\usepackage{bm}% bold math

\usepackage{psfrag}

\begin{document}

%\preprint{APS/123-QED}

\title{Optical absorption by magnesia-supported
gold clusters and nanocatalysts:
effects from the support, cluster and adsorbants}

\author{Michael Walter}
\author{Hannu H\"akkinen}
\email{hannu.hakkinen@phys.jyu.fi}

\affiliation{%
Department of Physics, Nanoscience Center, 
FIN-40014 University of Jyv{\"a}skyl{\"a}, Finland\\
}

\date{\today}% It is always \today, today,

\begin{abstract}
Polarization-resolved optical spectra
of magnesia-supported gold
clusters Au$_N$/MgO (N=1,2,4,8),
bound at a surface color center $F_s$ of the MgO(100) face, 
are calculated from the time-dependent density functional
theory. The optical lines for $N$=1,2
are dominated by 
transitions that involve
strong hybridization between gold
and  $F_s$ states 
whereas for $N$=4,8 
intracluster transitions dominate. The theoretical optical
spectra are  sensitive to cluster structure and adsorbants
(here CO and O$_2$ molecules on Au$_8$/$F_s$@MgO) 
which suggests 
polarization-resolved optical spectroscopy as
a powerful tool to investigate  structures and
functions of chemically active, supported clusters.
\end{abstract}

\maketitle

\section{Introduction}

 Investigations of the physical and chemical properties of metal clusters 
are currently largely motivated by the question how their various
remarkably size-dependent 
properties could be best utilized while 
{\it the clusters are interacting with the environment}, 
e.g., bound on or implanted in a support, or stabilized 
and surface-passivated by molecules \cite{meiwes}.
Understanding factors that dictate the
stability, structure and function has
 relevance regarding atomic-scale design 
of components that could be of potential use in future nanotechnologies. 
To this end, spectroscopic tools and density functional
theory (DFT) calculations can provide valuable insights.

Gold clusters and nanoparticles have attracted much attention
recently due to their remarkable chemical and optical
properties (for recent reviews,
see e.g. \cite{haruta,Pyykko:04}).  
Gold clusters with a size of just a few atoms,
supported by thin MgO films,
have turned out to be paradigms to provide insight into
the importance of {\it quantum finite size effects} in nanoscale
chemistry --  it has been
demonstrated that by changing the size or elemental
composition  of the supported cluster even
 {\it atom by atom} one can quite dramatically affect the catalytic properties
of the cluster \cite{hannu99,hannu03}. 
Earlier DFT calculations have shown that  
the finite-size effects in the electronic spectrum of the
nanocatalyst play a key role in CO oxidation reaction
through binding and activating the
oxygen molecule via a charge-transfer mechanism. Theoretical
predictions \cite{hannu99,hannu03} of charging of the catalytically active 
Au$_8$ cluster, bound at the surface color center  $F_s$
of the MgO film, were confirmed very recently by FTIR spectroscopy
\cite{yoon05}. 

Spectroscopic data on  structure 
 of supported metal clusters and ultrafine
nanoparticles are  still relatively scarce, with some notable exceptions
from STM and STS spectroscopy \cite{Harbich:01,Nilius:02}.
Optical spectroscopy is currently  a largely unexplored area, although
it has long been   used to study metal clusters in gas phase
\cite{Kreibig:95}.
Here we present a systematic theoretical investigation 
of the optical properties of
MgO-supported gold clusters, Au$_N$/MgO with $N\leq 8$, 
by analyzing the optical spectra
calculated from the linear response time-depenent DFT (TDDFT). We discuss
(i) the effects from the support, particularly from the $F_s$
states, dominating the spectra of $N=1,2$,
(ii) the sensitivity of the spectra to the atomic structure of the cluster
with examples for $N=4,8$, and finally 
(iii) the sensitivity of the spectra to adsorbants, by taking the
example of CO and O$_2$ adsorption on 
the smallest CO-oxidizing  cluster Au$_8$/MgO \cite{hannu99,hannu03,yoon05}.
It is shown that for all these cases, 
{\it polarization-resolved} spectra contain rich information on the nature of
the dominant optical transitions and their connection to the structure and
shape of the adsorbed cluster.

\section{Ground-state DFT calculations }

The atomic and electronic structure 
of the supported Au$_N$/MgO system 
(comprising  the vicinity of the color center 
of the MgO(100) surface, the adsorbed Au$_N$ 
cluster, and O$_2$ and CO molecules) 
were calculated within the DFT 
in combination with Born-Oppenheimer (BO) 
Molecular Dynamics (MD)\cite{Barnett:93}
including self-consistent 
gradient corrections via the so-called PBE-GGA 
functional \cite{PBE:96}.
 Au($5d^{10}6s^1$), Mg($3s^2$), C($2s^22p^2$), and O($2s^22p^4$)
 electrons were included in the valence, 
and the interaction to the  ion cores was 
described by scalar-relativistic non-local 
norm conserving pseudopotentials devised by 
Troullier and Martins \cite{Troullier:91,fn1}.

The MgO surface  
is modelled by a two-layer ab initio cluster Mg$_m$O$_m$ or $F_s$@Mg$_m$O$_{m-1}$, 
embedded in an extended point-charge lattice to include effects of
the long-rage Madelung potential \cite{pacch}.
For the embedding lattice,
around 2100 alternating charges of +2 and -2 representing
Mg and O ions, respectively, were used. 
In addition, those positive point charges 
that would be nearest neighbors 
to the peripheral O atoms of the central ab initio Mg$_m$O$_n$
cluster have been replaced by "empty"  
Mg pseudopotentials (MgPP) in order to prevent unphysical 
polarization effects to O ions 
(see visualization of Au/Mg$_{13}$O$_{13}$
in Fig. 1a)
\cite{pacch}.
The lattice parameter of the embedding part 
is fixed to the experimental lattice constant 
(4.21 \AA) of bulk MgO. The cluster, molecules
and nearest-neighbor Mg ions to the $F_s$ are treated
dynamically in structural optimizations that included
both steepest-descent and quenched molecular dynamics
runs \cite{hannu03,fn3}.

\section{  Linear-response TDDFT calculations} 

We follow the formulation of the linear response time dependent
DFT given by Casida, \cite{Casida:96}
as implemented in ref. \cite{mos01}. Briefly,
in order to get the weights $F_I$ and energies $\hbar\omega_I$ of
optical transitions $\lbrace I\rbrace$, one solves an eigenvalue problem 
$\Omega F_I = \omega^2_IF_I$
where the $\Omega$ matrix elements are given by
$$\Omega_{ij,kl}=\delta_{ik}\delta_{jl}\varepsilon^2_{ij} +
2\sqrt{n_{ij}\varepsilon_{ij}n_{kl}\varepsilon_{kl}} K_{ij,kl} .
\eqno(1)$$
Here $\varepsilon_{ij}=\varepsilon_j - \varepsilon_i$ and
$n_{ij}=n_i-n_j$ are the 
difference of the KS particle-hole eigenvalues
and occupation numbers, respectively.
$K_{ij,kl}$ is a coupling matrix that describes the
linear response of the electron density $\rho$
to the single-particle
 -- single-hole excitations in the basis spanned by the 
ground state KS orbitals
$\vert i\rangle$.
The transition matrix element in polarisation 
direction $\hat\epsilon_{\nu}$ is
$$ (M_I)_{\nu} = \sum_{ij}\sqrt{\varepsilon_{ij}n_{ij}}
\langle j\vert r_{\nu}\vert i\rangle(F_I)_{ij} \eqno(2)$$
and the corresponding polarization-dependent
and polarization-averaged oscillator strengths are
$(f_I)_{\nu}=2\vert(M_I)_{\nu}\vert^2$
and $\bar f_I =(1/3)\sum_{\nu=1}^3(f_I)_{\nu} $, respectively.
The convergence of the spectra (oscillator strengths and
energies of the major peaks) depends basically on 
three factors: (i) extent of the KS particle-hole basis,
(ii) size of the plane-wave grid,  
and (iii) size of the ab initio Mg$_m$O$_n$ cluster. Factors (i) and
(ii) determine the quality of 
discretization of the continuum virtual states
while factor (iii) dictates the overall behavior in the optical
response of the MgO support - specifically the width of the band
gap. We have carefully tested the convergence behavior of the
calculated spectra in case of the smallest supported clusters
($N$=1,2) by using Mg$_m$O$_{m-1}$ clusters extending up to $m=49$.
The converged theoretical optical band gap of the support is about 4
eV, which is in a fair agreement with the apparent gap of 6 eV
measured for a few-layer thick MgO film \cite{schintke}, taking into account
the general tendency of DFT to underestimate band gaps of wide-gap
insulators. However, the important transitions induced by the
supported cluster fall into the band gap. These transitions can be
calculated within 0.1 eV accuracy even with significantly smaller
cluster sizes ($m=13,25$).

\begin{figure}[t]
  a)
  \parbox{6cm}{
    \includegraphics[scale=0.3]{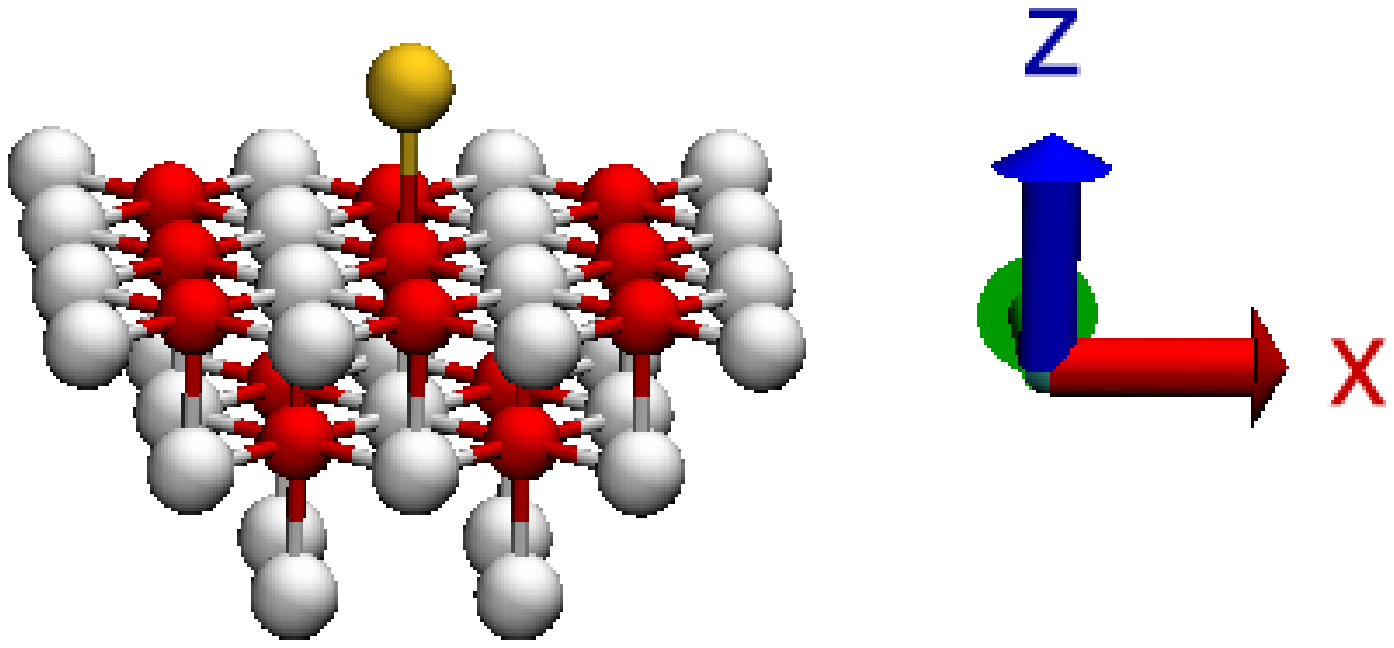}% Here is how to import
    % EPS art
  }
  \\
  \psfrag{b}{b)}
  \psfrag{c}{c)}
  \psfrag{d}{d)}
  \psfrag{e}{e)}
  \psfrag{omega}{$\hbar\omega$ [eV]}
  \psfrag{omic}{{\footnotesize $\hbar\omega$ [eV]}}
  \psfrag{lambda}{$\lambda$ [nm]}
  \includegraphics[scale=0.33]{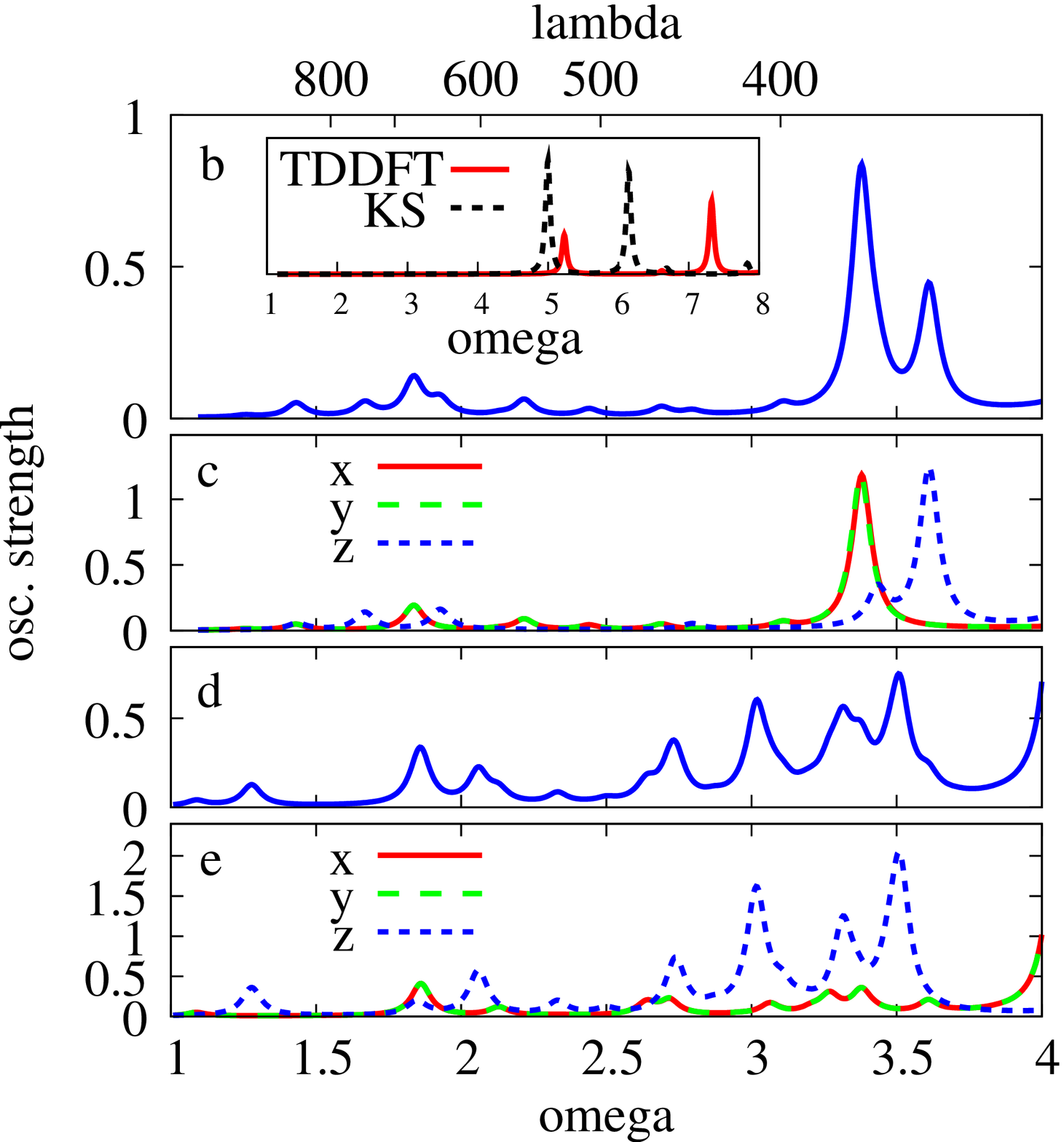}% Here is how to import EPS art
  \caption{\label{fig:Au1} (color online)
    The supported Au atom: a) The ab initio cluster
    Au/Mg$_{13}$O$_{13}$ with the peripheral "empty" MgPP sites, b) the
    polarization averaged spectrum (the inset shows the spectrum of a gas
    phase Au atom), c) the polarization resolved
    spectrum. d) and e) are polarization averaged and resolved spectra
    for the atom on $F_s$. All spectra here and in subsequent figures are plotted by
using a Lorenzian folding of 0.08 eV of individual optical lines.
Note, that the spectra in d,e have been calculated for
Au/$F_s$@Mg$_{25}$O$_{24}$. }
\end{figure}

\section{Results and Discussion}

Application of the TDLDA method to  an isolated gold atom leads
to the optical spectrum shown in the inset to Fig. 1b (for comparison,
we show also the optical spectrum based on KS-only transitions,  i.e.,
by setting $K_{ij,kl}=0$ in Eq. (1)).
This spectrum was calculated by considering the 
(5d$^{10}6s^16p^07s^07p^0$) KS states in the particle-hole basis.
The KS spectrum consists of two strong transitions corresponding to
single-particle excitations of $6s\rightarrow 6p$ (5 eV) and
$5d\rightarrow 6p$ (6.1 eV).
The lowest TDLDA optical line is found at 5.23 eV and it is
an incoherent (destructive)
superposition of the KS transitions $6s\rightarrow 6p$ 
and $5d\rightarrow 6p$ (the destructive interference is manifested
by the reduced oscillator strength of the 5.23 eV TDLDA line
compared to the neighboring KS lines). Our result of
5.23 eV is in a fair agreement with  the lowest
experimental  spin-orbit (SO) split 
transitions   at 4.63 eV ($^2S_{1/2} \rightarrow\
^2P_{1/2}$, final single-particle state assigned as ($5d^{10}6p^1$))
 and 5.11 eV ($^2S_{1/2}\rightarrow\
^2P_{3/2}\ \ (5d^{10}6p^1)$) \cite{NIST} taking into account the fact
that the SO effect is not included in our calculations.  
Similarly, the next strong TDLDA transition at 7.3 eV
corresponds to the experimental SO doublet 7.44 eV
and 7.53 eV,  assigned to transitions to the
$^2P_{1/2},\ ^2P_{3/2}$ states with a ($5d^{10}7p^1)$ single-particle
configuration \cite{NIST}. We also note here, that another recent 
TDDFT calculation \cite{Antonietti:05} finds the first optical
transition of an isolated Au atom at 5.23 eV, in full agreement with
our result \cite{fn4}.

For the combined Au/MgO system,
we find the lowest strong transitions
at 3.38 eV and 3.61 eV, i.e,  in the
near-UV region (see  Fig. 1b).
The nature of these lines is
revealed by the 
the polarization-resolved spectra (Fig. 1c)
which shows basically three peaks that have degenerate $(f_I)_\nu$ 
values, two of which (x,y polarization)
 at 3.38 eV and the z-polarized transition (i.e., excitable for
z-polarized light) at 3.61 eV.
The identical oscillator strengths in the x,y,z directions and
the degeneracy pattern strongly suggest that these transitions
mainly originate from the Au atom, perturbed and  shifted by the support.
(Note that neither the ideal MgO surface nor the
gas phase atom have transitions anywhere near 3.5 eV).
One can also locate a set of weak transitions in the region
1.7 eV -- 1.9 eV that consists of three weak lines, of which two
are z-polarized and one xy-polarized. 
In general, these lines are due to incoherent mixing of several
KS transitions, but we note that they have a component of
the atomic $5d\rightarrow 6s$ character, which is dipole-forbidden
in the gas phase atom but becomes partially "visible" via mixing with
the substrate states.

The situation is dramatically changed when the $F_s$ defect
is introduced (Fig. 1d,e). Several strong lines are now
seen in the range 1.8 eV -- 3.5 eV. Fig. 1e reveals strikingly that
basically {\it all} of the transitions are polarized only
in the z-direction, i.e., perpendicular to the surface. A stronger coupling
to the $F_s$ states also enhances the oscillator strength of the bands
around 2 eV which have the contribution from the gas-phase forbidden
 $5d\rightarrow 6s$ transition.

\begin{figure}[t]
  a)
  \parbox{6cm}{
    \includegraphics[scale=0.20]{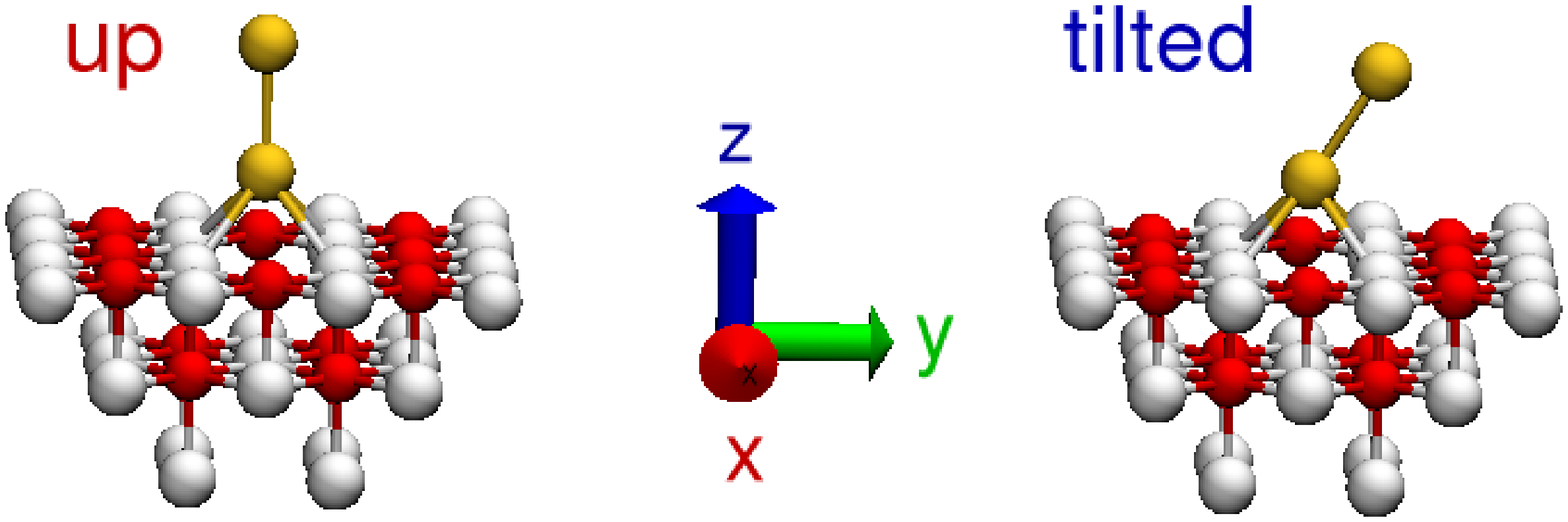}% Here is how to import
    % EPS art
  }
  \\
  \psfrag{b}{b)}
  \psfrag{c}{c)}
  \psfrag{d}{d)}
  \psfrag{omega}{$\hbar\omega$ [eV]}
  \includegraphics[scale=0.33]{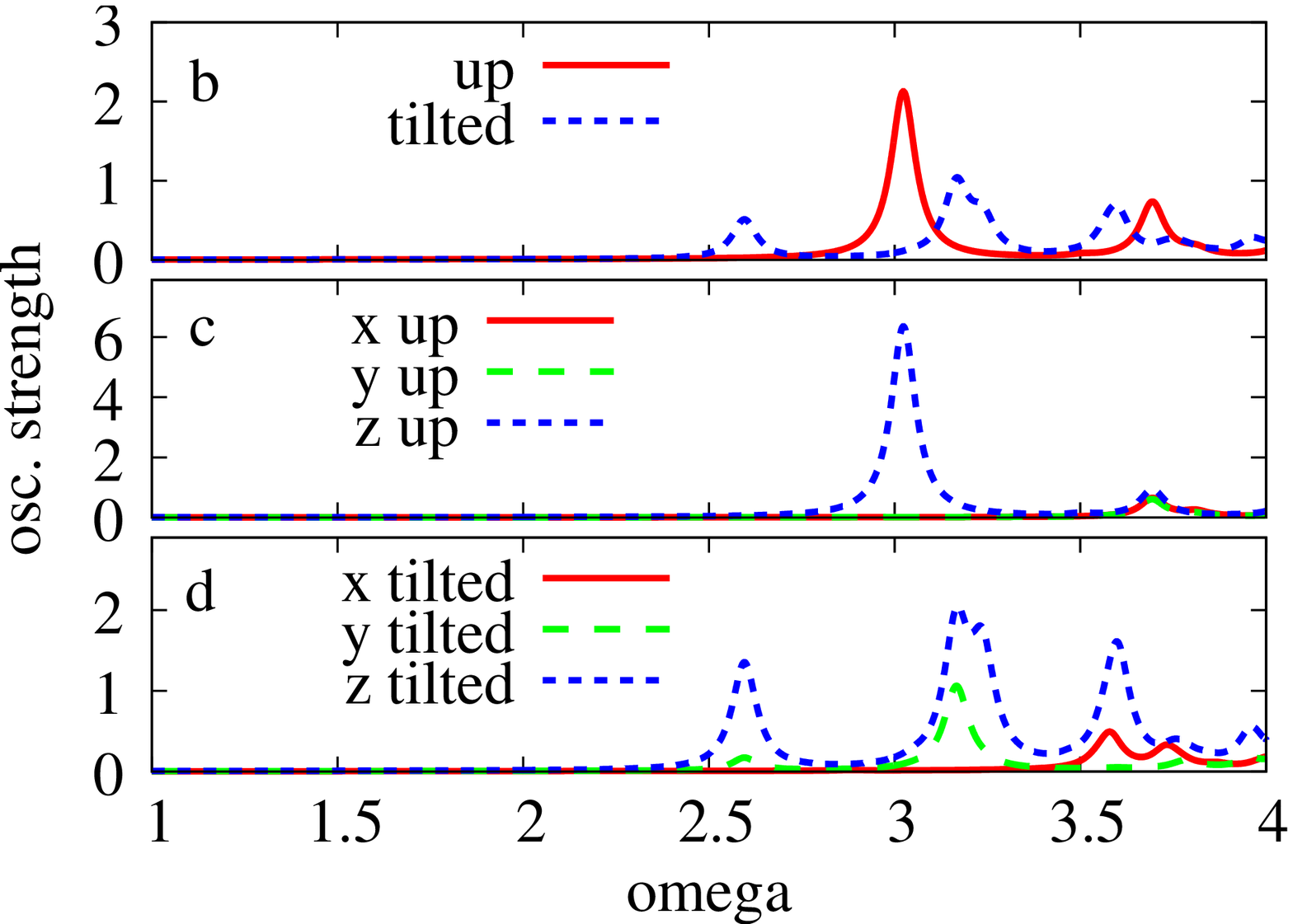}% Here is how to import EPS art
  \caption{\label{fig:Au2} (color online)
    a) Two configurations for Au$_2$ bound to $F_s$, b) polarization
    averaged and c,d) resolved spectra. $d$(Au-Au)=2.61 \AA\ and 2.62 \AA\
    for "up" and "tilted" geometries, respectively,
    compared to
    the gas-phase value of 2.55 \AA.
  }
\end{figure}
For Au$_2$, bound at $F_s$, we have found  two almost isoenergetic
geometries  (Fig. 2a); the "tilted" geometry with the Au-Au axis $33^o$
from the surface normal is 0.1 eV more stable than the dimer standing
"up". This slight difference in the adsorption geometry shows up
drastically in the TDLDA spectrum (Figs. 2b-d): the "up" geometry
has transitions only in the z-direction whereas the spectrum of the
"tilted" cluster has a clear y-component at 3.1 eV
(y is the direction to which the dimer is tilted).
Based on the clear difference between the TDLDA spectra
in Figs. 2b-d we predict  that in this simplest non-trivial case of 
adsorption of a gold cluster on the defect site of MgO surface, experimental
optical spectra could yield a definitive answer on the binding geometry.

\begin{figure}[t]
  \psfrag{a}{a)}
  \includegraphics[scale=0.18]{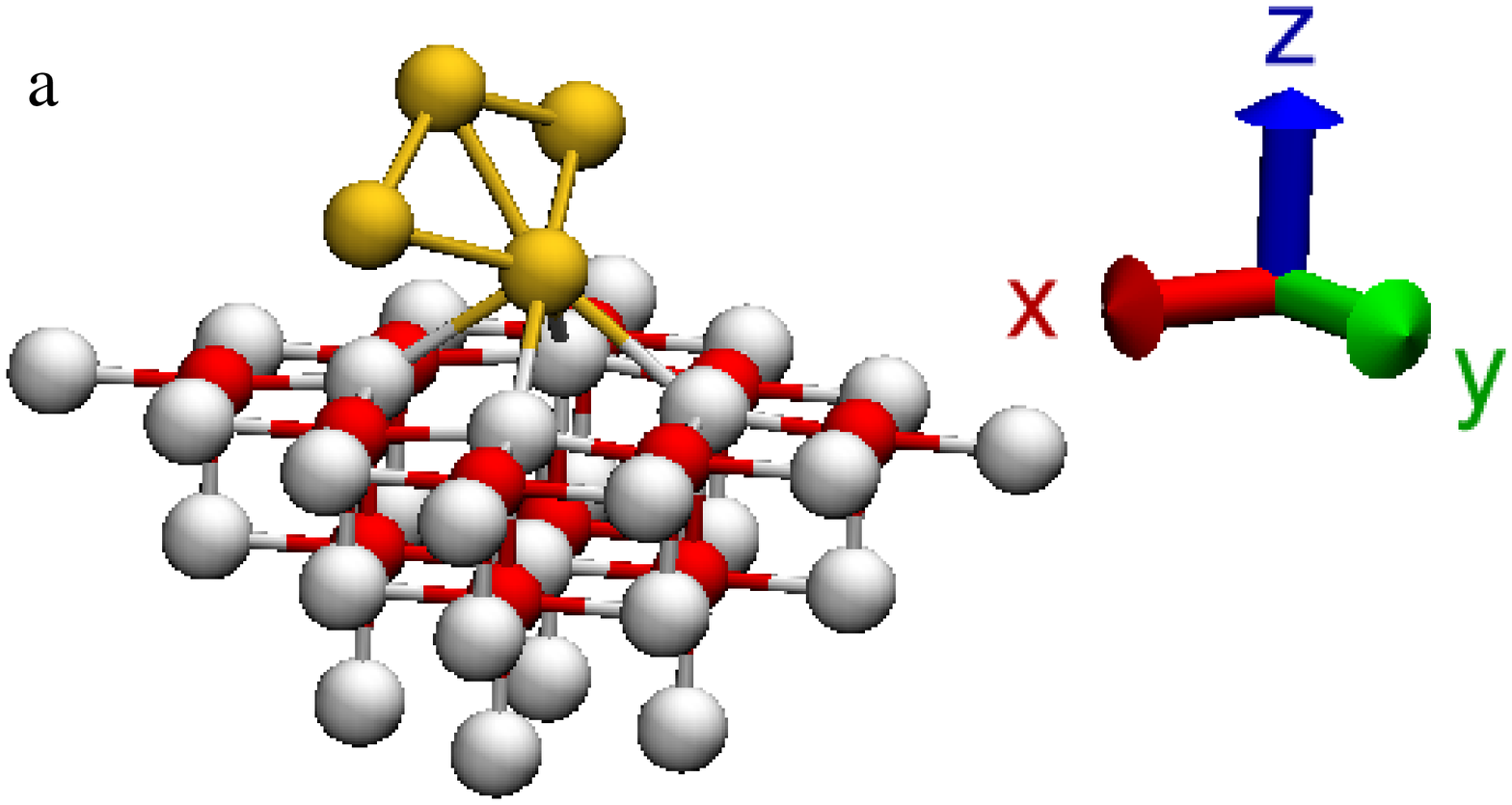}% Here is how to import
  \psfrag{b}{b)}
  \psfrag{omega}{$\hbar\omega$ [eV]}
  \includegraphics[scale=0.3]{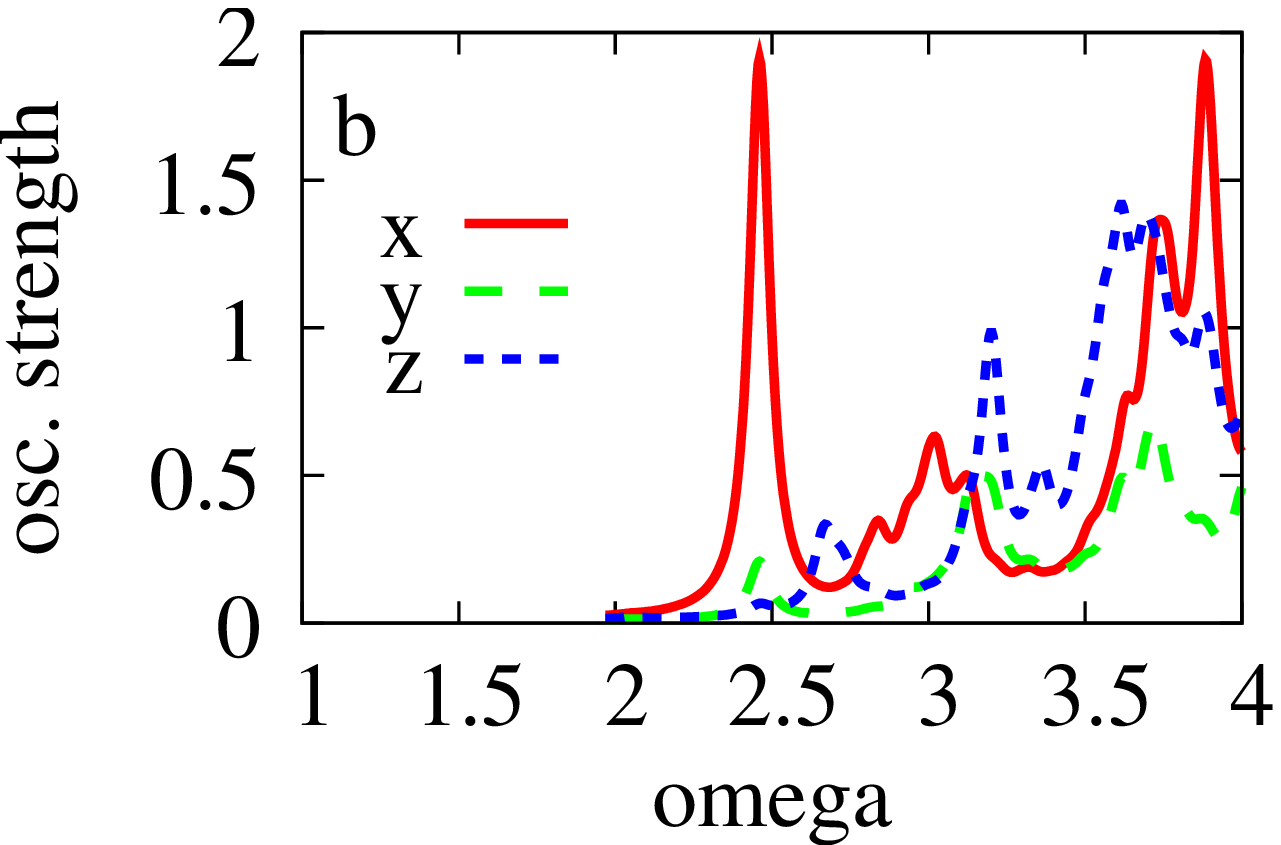}% Here is how to import EPS art
  \caption{\label{fig:Au4} (color online)
    The configuration a) and polarization resolved spectrum b) of
    Au$_4$ bound to $F_s$.
  }
\end{figure}
Au$_4$, whose  optimum binding geometry has previously
been found to be that of a tilted rhombus-like structure\cite{hannu03} (Fig. 3a),
has a very strong transition close to 2.5 eV (Fig. 3b). This transition
is solely in the direction of the long axis of the cluster. The tetramer
is thus the smallest cluster which "screens out" the
transitions involving major (z-)components from the color center, which
had been found to be dominant for the monomer and dimer.

\begin{figure}[t]
  a)
  \parbox{6cm}{
    \includegraphics[scale=0.18]{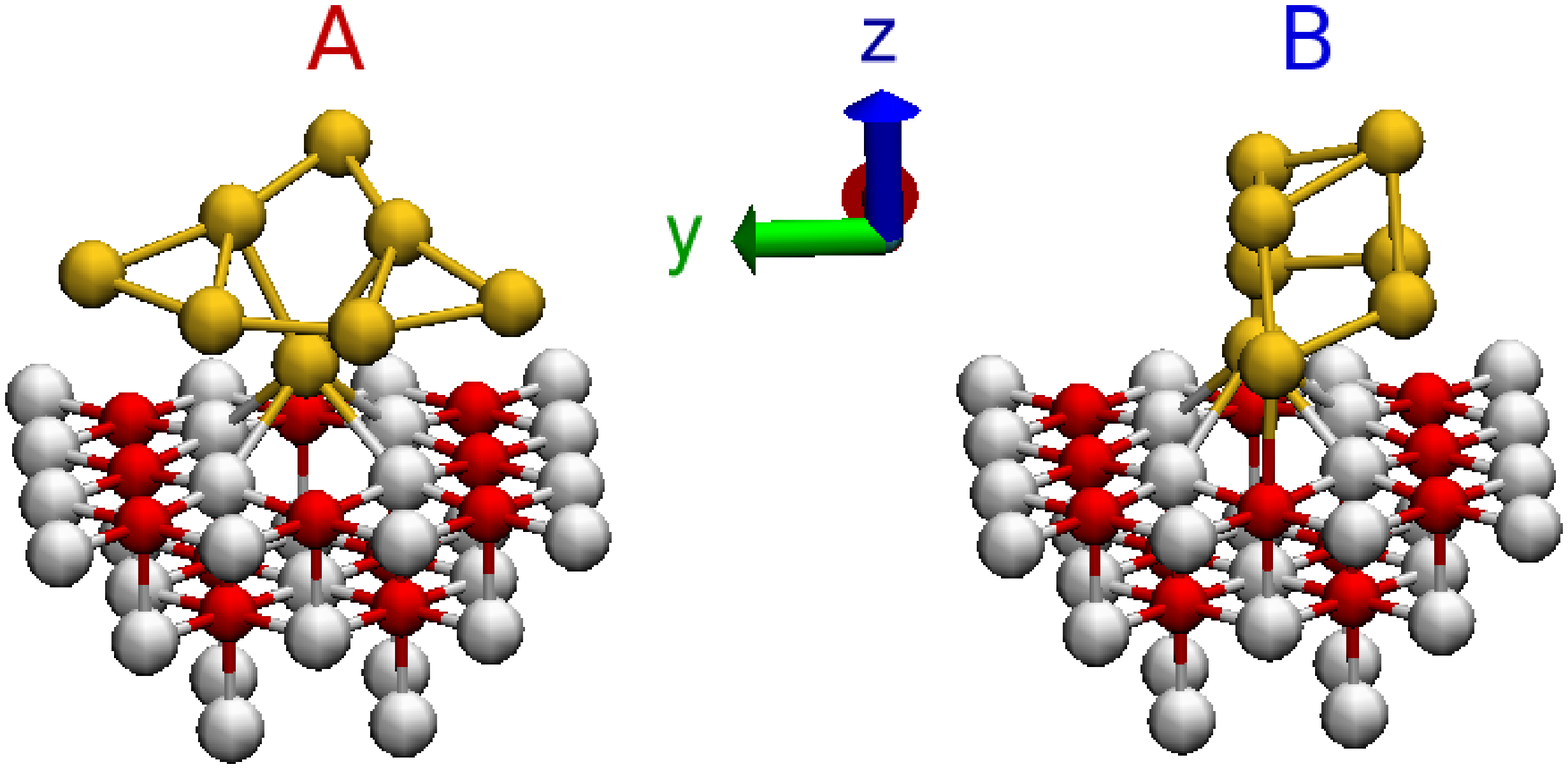}% Here is how to import
    % EPS art
  }
  \\
  \psfrag{b}{b)}
  \psfrag{c}{c)}
  \psfrag{omega}{$\hbar\omega$ [eV]}
  \includegraphics[scale=0.33]{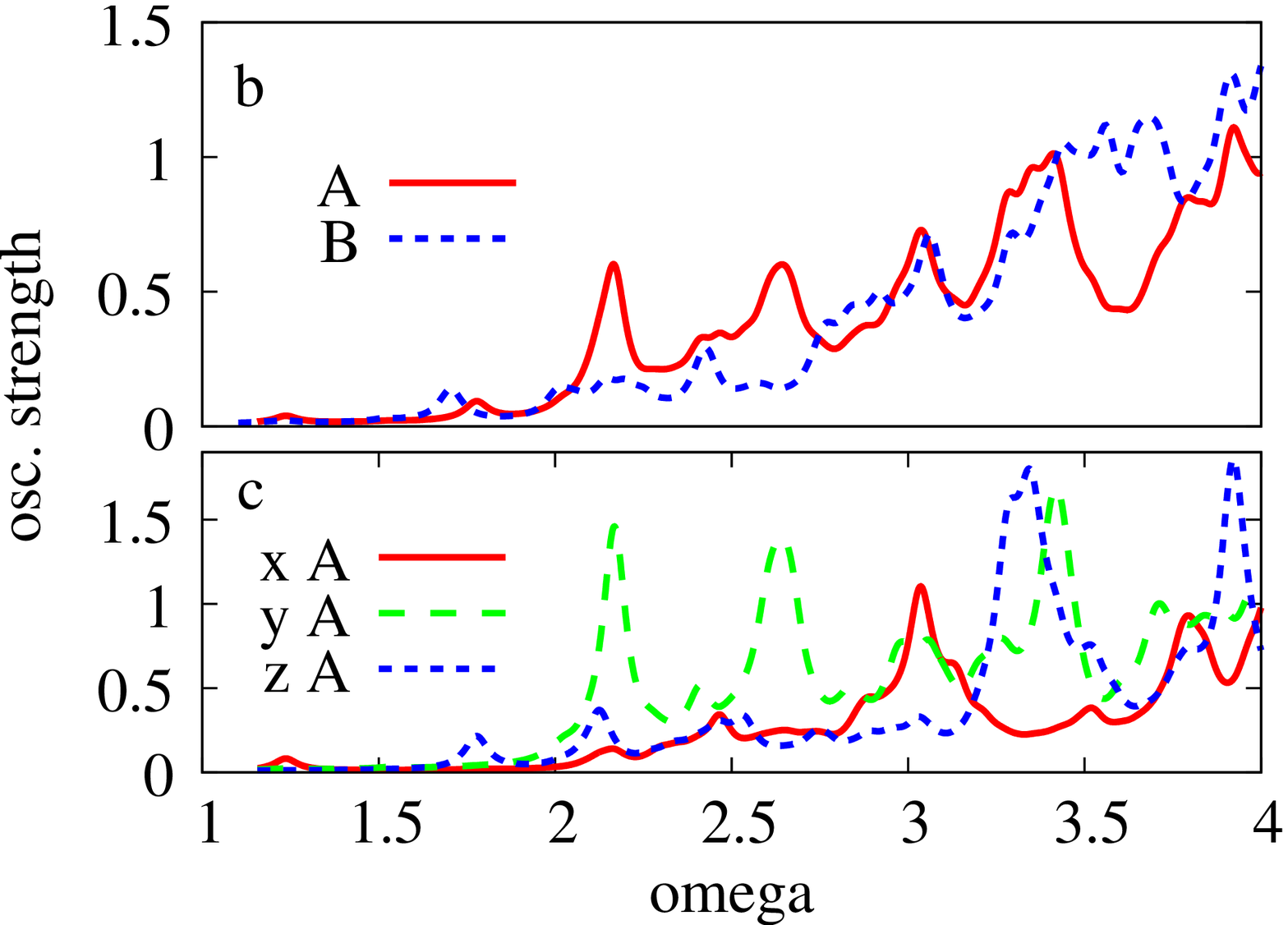}% Here is how to
                                % import EPS art
  \caption{\label{fig:Au8} (color online)
    Au$_8$ on $F_s$: a) Two configurations and b) their polarization
    averaged spectra. c) The polarisation resolved spectra of
    structure A.
  }
\end{figure}
\begin{figure}[t]
  a)
  \parbox{6cm}{
    \includegraphics[scale=0.19]{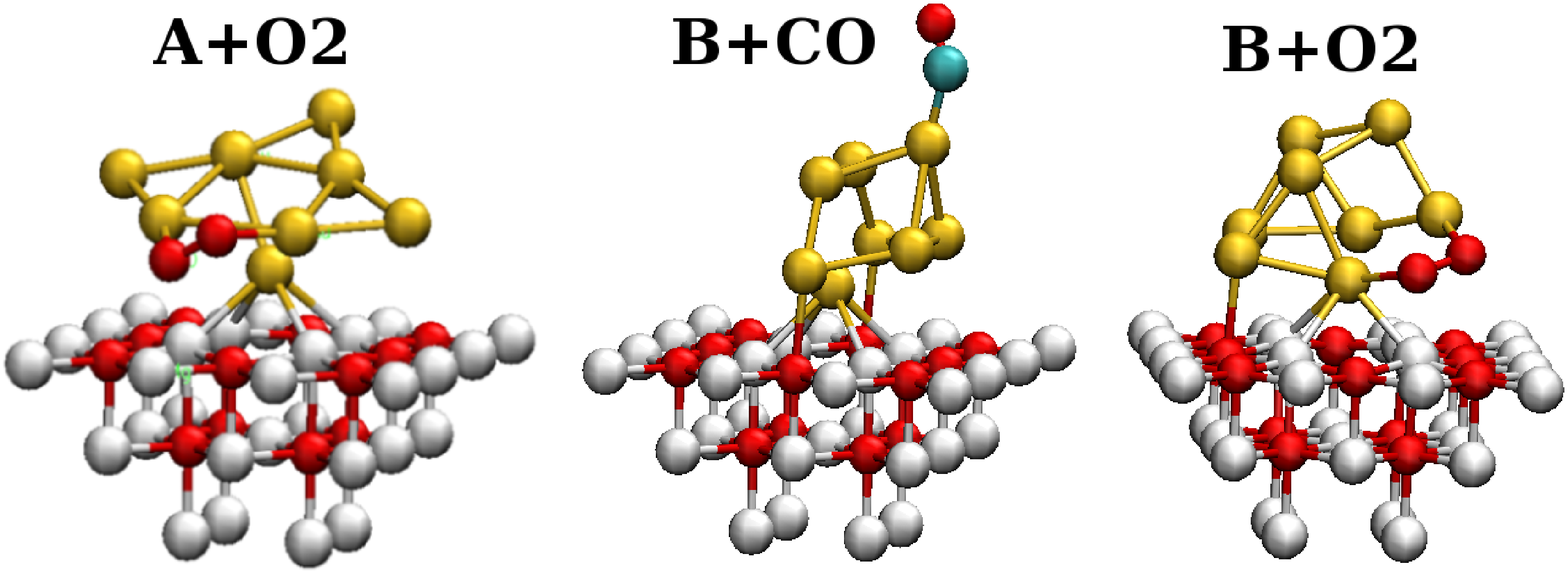}% Here is how to import
    % EPS art
  }
  \\
  \psfrag{e}{b)}
  \psfrag{f}{c)}
  \psfrag{omega}{$\hbar\omega$ [eV]}
  \includegraphics[scale=0.33]{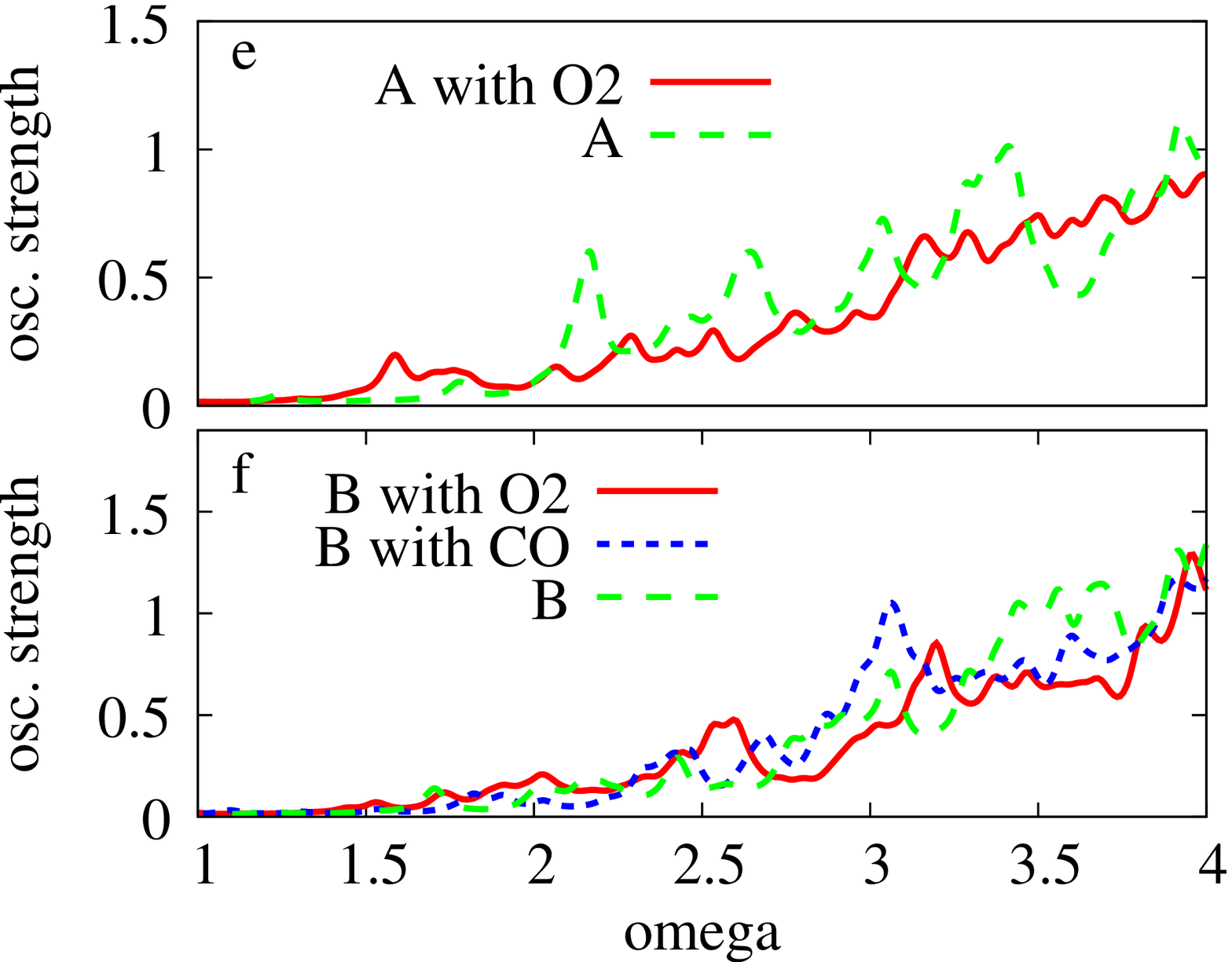}% Here is how to
                                % import EPS art
  \caption{\label{fig:Au8} (color online)
    Au$_8$ and absorbed molecules on $F_s$: 
    a) The configurations of the
    adsorbed molecules and the spectra of b) structure A with and
    without O$_2$ and c) structure B with and without CO/O$_2$. 
  }
\end{figure}
We now turn to discussion of the optical properties of the
"smallest MgO-supported nanocatalyst" for
CO oxidation, namely Au$_8$. \cite{hannu99,hannu03,yoon05} 
Two isomeric structures were discussed in Ref. \cite{hannu03}, where
the open structure A  is more stable than the compact
structure B by 0.3 eV (see Fig. 4a, note that similar geometric motifs
yield locally stable isomers also for neutral gas-phase gold octamers
where open structures are favored due to relativistic bonding effects
\cite{hannuwang,henrik}).  Fig. 4b shows that these two isomers yield
very different, characteristic spectra: while the more stable open structure
A has a series of intense transitions covering most of the optical range
and yielding characteristic line shapes, the spectrum of structure B has
less oscillator strength below 2.7 eV. 
 Fig. 4c gives more insight into the strong
absorption bands at 2.1 eV, 2.4 eV,
 and 2.65 eV of structure A: they are mostly
transitions involving the long axis ($\sim$y-direction)
 of the cluster. Bands after
3 eV start to have major contributions also from transitions in the other
two polarization directions. 
Again, Figs. 4b-c suggests that in principle it should be possible to
distinguish different isomers of adsorbed gold octamers
via high-resolution optical data.

\begin{table}[htp]
  \begin{tabular}{l | r | r | r | r}
    Structure & Au$_8$ & CO & O$_2$ & support \\
    \hline
    A & -0.37 & & & +0.37 \\
    A+O$_2$ & +0.73 & & -1.26 & +0.53 \\
    B & -0.20 & & & +0.20 \\
    B+O$_2$ & +0.60 & & -1.11 & +0.51 \\
    B+CO & +0.38 & -0.51 & & +0.13
  \end{tabular}
  \caption{\label{tabel1}
    Local charges\cite{fn2} (in units of $e$) on the constituents of
    the structures shown in Figs. 4a) and 5a). Support includes the
    MgO surface with the color center.
  }
\end{table}

Finally, we show examples about modification of the optical
spectra by adsorbant molecules in Fig.\ 5, here CO and O$_2$.
Fig. 5b shows that adsorption of the O$_2$ molecule to the structure A
of the gold octamer strongly suppresses the absorption bands at 2.1 eV,
2.4 eV, and 2.65 eV, furthermore, a new clear absorption band appears
now at 1.5 -- 1.7 eV. Analysis of the polarization-resolved spectra
(not displayed) shows  disappearance of the  bands
connected to excitations along the long axis of structure A. 
On the other hand, adsorption of CO or O$_2$ to the structure B of Au$_8$
does not significantly modify the absorption spectrum below 3 eV
(Fig. 5c). 
This finding 
can be correlated to {\it charge transfer} from Au$_8$ to O$_2$ (see
Table 1): local
charge on the MgO-supported gold octamer (structure A)
changes from $-0.37 e$ to $+0.73 e$
upon oxygen adsorption, and O$_2$ gains a charge of $-1.26e$ that goes
to the originally empty $2\pi^*$ antibonding orbitals, i.e., 
the adsorbed dioxygen is activated to a peroxo-like state.\cite{fn2}
This charge-transfer-induced activation
 was shown to be the key to the catalytic acitivity.
\cite{hannu99,hannu03,yoon05}
In the TDDFT framework, the significant change of the oxidation  state
of  the gold octamer A
upon dioxygen adsorption ($1.1e$ in total)
 obviously depletes those single-particle KS states
from the TDDFT basis, that have a major contribution to the strong
excitations of the bare cluster in the energy range shown.
O$_2$ and CO  adsorption on structure B are accompanied by
a  smaller change of the oxidation state of gold  ($0.8e$
and $0.58e$ in total, respectively).

We have shown that polarization-resolved optical spectra
of magnesia-supported gold clusters contain rich information about
the binding mode and structure of the cluster and possible adsorbant
molecules. Specifically,  optical spectra contain complementary
information about the adsorption of CO and O$_2$ molecules
on the catalytically active center Au$_8$/$F_s$@MgO and the related 
charge transfer.
 Emerging applications of surface-sensitive optical spectroscopic
tools \cite{Antonietti:05}
 to this and other chemically active supported metal clusters
are likely to help significantly in gaining detailed understanding
of structures and functions of reaction centers of nanocatalysts.

It is a pleasure to thank  M. Moseler for numerous discussions
on the TDDFT method.  Computations were
performed on IBM SP4  at the  CSC -- the Finnish IT Center for
Science in Espoo.
This work was partially supported by 
the Academy of Finland (AF). HH acknowledges a bilateral AF -- DAAD
travel grant on project "Supported Metal Clusters and Nanoparticles:
Electronic Structure, Optical Properties and Nanocatalysis".

%\cite{Schintke01prl}
%Hakkinen03ac,
%Yoon05sc,
%Schaub01prl,
%Nilius02sc,
%Perdew96prl,
%Troullier91prb,
%Moseler01prl,
%Hakkinen03jpca,
%Antonietti05prl
%}

%\input{tables}

%\bibliography{/home/miwalter/tex/util/all}
%\bibliography{biblio}% Produces the bibliography via BibTeX.

\begin{thebibliography}{23}
\bibitem{fn1}{We use  the core radii 
(in $a_0$, "tilde" denotes the local component)
 of: Au: $\tilde s$(2.5), $p$(3.0), $d$(2.0); 
O: $s$(1.45), $\tilde p$(1.45); C: $s$(1.50), 
$\tilde p$(1.54); Mg: $s$(2.5), $\tilde p$(2.75). 
 We expand the KS states in a plane wave basis with 62 Ry cutoff.}
\bibitem{fn3}{
The structures 
for Au$_N$/$F_s$@MgO with $N=1,4,8$ have been reported previously in
ref. \cite{hannu03}, in addition, here we optimized the structures
for $N=2$. 
See Refs.\ \cite{hannu99} and \cite{pacch} for a general discussion on
the suitability of embedded-cluster DFT models for description of
oxide-supported metal clusters.
}
\end{thebibliography}

%\begin{figure*}
%\includegraphics[width=8cm]{}% Here is how to import EPS art
%\caption{\label{fig:1} fig caption 1  }
%\end{figure*}

%\begin{figure*}
%\includegraphics[width=8cm]{}% Here is how to import EPS art
%\caption{\label{fig:2} fig caption 2  }
%\end{figure*}

%\begin{figure*}
%\includegraphics[width=8cm]{}% Here is how to import EPS art
%\caption{\label{fig:3} fig caption 3 }
%\end{figure*} 

%\begin{figure*}
%\includegraphics[width=8cm]{}% Here is how to import EPS art
%\caption{\label{fig:4} fig caption 4 }
%\end{figure*}

\end{document}